\documentstyle[psfig,aps,prl,amssymb]{revtex}

\begin{document}
\title{Monotonicity Properties of Certain Measures over 
the Two-Level Quantum Systems}
\author{Paul B. Slater}
\address{ISBER, University of
California, Santa Barbara, CA 93106-2150\\
e-mail: slater@itp.ucsb.edu,
FAX: (805) 893-7995}

\date{\today}

\draft
\maketitle
\vskip -0.1cm

\begin{abstract}
We demonstrate --- using the case of the two-dimensional quantum
systems --- that the ``natural measure on the space of density matrices
$\rho$ describing $N$-dimensional quantum systems'' proposed by
\.Zyczkowski {\it et al} 
[Phys. Rev. A \textbf{58}, 883 (1998)] does {\it not} 
belong to the class of
normalized volume
elements of monotone metrics on the quantum systems.
Such metrics  possess the statistically important  property
of being decreasing under stochastic mappings (coarse-grainings).
We do note that the proposed natural measure
(and certain evident variations upon it) exhibit quite specific
monotonicity characteristics, but not of the form
required for membership in that distinguished class.

\end{abstract}

\vspace{.2cm}
\hspace{1.5cm} Keywords: density matrix, monotone metric, operator monotone function, Bures metric,

\hspace{1.5cm} measures over quantum systems

\vspace{.15cm}

\hspace{1.5cm} Mathematics Subject  Classification (2000): 81Qxx, 26A48


\vspace{.1cm}

In a recent paper \cite{zycz1},
 \.Zyczkowski, Horodecki, Sanpera and Lewenstein (ZHSL)
 proposed a ``natural measure in the space of density matrices
$\rho$ describing $N$-dimensional quantum systems''.
We demonstrate here --- using the two-dimensional quantum
 systems --- that this measure (and certain obvious variations upon
it) do not belong to another  class of, arguably,   
``natural measures'' (ones consistent with desired properties under
``coarse-graining'' \cite{quark,calz}). This collection of
 measures, recently studied by Slater
in various contexts
\cite{slater2,slaterapriori,slatertherm} (cf. \cite{kratt}), consists of the 
normalized volume elements of monotone metrics \cite{petzsudar}
(including, notably, 
 the {\it minimal} monotone or {\it Bures} metric \cite{hubner,brauncaves}). 
Nevertheless, the ZHSL measure and the variations 
we associate with it below ((\ref{cartesian}) and
(\ref{sph})), exhibit quite definite monotonicity
properties, but not those needed for these measures to fall within
the indicated class.

Using the ``Bloch sphere'' representation \cite{braun} of the two-dimensional
density matrices,
\begin{equation} \label{1}
\rho = {1 \over 2} \pmatrix{1 +z & x - i y \cr
x + i y & 1 - z\cr},
\end{equation}
where $x^2 +y^2 +z^2 \leq 1$, and converting to spherical coordinates
 $(x = r \sin{\theta} \cos{\phi}, y = r \sin{\theta} \cos{\phi}, z = r
\cos{\theta})$, an (unnormalized) volume element
of a monotone metric is expressible in the form 
(cf. \cite[eq. (3.17)]{petzsudar}),
\begin{equation} \label{monotone}
{r^2 \sin{\theta} \over  f \big((1-r)/(1+r)\big) (1-r^2)^{1/2} (1+r)}.
\end{equation}
Here $f(t)$  is an {\it operator monotone} function
satisfying a (self-adjointness) condition, $f(t) = t f(1/t)$. All such 
functions considered in \cite{petzsudar} also fulfill a normalization
condition (at least, in a limiting sense), $f(t) = 1$.
A function $f: {\Bbb{R}}^{+} \to {\Bbb{R}}$ is called operator monotone
if the relation $0 \leq K \leq H$ implies $0 \leq f(K) \leq f(H)$ for
any matrices $K$ and $H$ of any order. The relation $K \leq H$ signifies
that the eigenvalues of $H-K$ are non-negative. (Plenio, Virmani, and
Papadopoulos \cite{plenio} have recently employed operator monotone
functions to derive a new inequality relating the quantum relative
entropy and the quantum conditional entropy. For another quantum-theoretic
study involving such functions, cf. \cite{andy}.)

Now, using the polar decomposition theorem, ZHSL represented density matrices
in the form,
\begin{equation} \label{polar}
\rho = U  D U^{\dagger},
\end{equation}
where $U$  is a unitary matrix, $U^{\dagger}$ is its conjugate
transpose, and $D$ is a diagonal matrix
with non-negative elements $d_{i}$ (the eigenvalues
of $\rho$), which, of course,
sum to 1.
The action of the operator monotone function $f(t)$
 upon $\rho$ can be
expressed in the form
\begin{equation}
 f(\rho) = U f(D) U^{\dagger},
\end{equation}
where $f$ acts individually on the diagonal entries of $D$
 \cite[p. 112]{bhatia}.
The ZHSL measure --- which ZHSL 
 used for estimating the volume of separable states
\cite{zycz1} (cf. \cite{slaterapriori,zycz2}) --- is the
 product of the uniform distribution (Haar measure)
on the unitary transformations 
$U(N)$ and the uniform distribution on
the $(N-1)$-dimensional simplex spanned by the $N$ eigenvalues
($d_{i}$) of $\rho$.

It is quite natural to consider this latter uniform distribution as the
specific case ($\nu =1$) of the $(N-1)$-dimensional 
(symmetric) Dirichlet 
probability distributions \cite{wilks,ferguson}
 \cite[eq. (1)]{slaterapriori} \cite[eq. (3)]{zycz2},
\begin{equation} \label{dirichlet}
p_{\nu}(d_{1}, d_{2},\ldots,d_{N}) =
 {\Gamma(N \nu)  d_{1}^{\nu -1} d_{2}^{\nu -1}
\ldots d_{N}^{\nu -1} \over \Gamma(\nu)^N}, \quad \nu >  0
\end{equation}
on the $(N-1)$-dimensional simplex.
(For $\nu= {1 \over 2}$, one obtains the ``Jeffreys' prior'' of Bayesian
theory \cite{kass},
 corresponding to the classically {\it unique} monotone/Fisher
information metric \cite{chentsov}.
\.Zyczkowski \cite[App. A]{zycz2} has shown that a vector of an $N$-dimensional
random orthogonal [unitary] matrix generates the Dirichlet measure
(\ref{dirichlet}) with $\nu = {1 \over 2}$ $[\nu = 1]$.)

For $N =2$, the unitary matrices $U$ are parameterizable
 as the product of a phase factor (irrelevant for
the purposes here) and a member of $SU(2)$ \cite[eqs. (2.40), (2.41)]{bied},
\begin{equation}
U(\alpha \beta \gamma) = e^{-i \alpha \sigma_{3}/2} e^{-i \beta \sigma_{2}/2}
 e^{-i \gamma \sigma_{3}/2},
\end{equation}
where $\sigma_{i}$ denotes the $i$-th Pauli matrix and the three Euler
angles have the ranges, 
$0 \leq \alpha < 2 \pi$, $0 \leq \beta \leq \pi$, and $0 \leq \gamma < 2 \pi$.
More explicitly, we have 
\begin{equation} 
 U(\alpha \beta \gamma) =
 \pmatrix{ e^{-i \alpha / 2} \cos(\beta / 2) e^{-i \gamma / 2}
 &  -e^{-i \alpha / 2} \sin(\beta / 2) e^{i \gamma / 2} \cr
e^{i \alpha / 2} \sin(\beta / 2) e^{-i \gamma / 2}
 & e^{i \alpha / 2} \cos(\beta / 2) e^{i \gamma / 2} \cr}.
\end{equation}
Since the  angle $\gamma$ also can be shown to be absent 
(drop out) in the ZHSL representation
(\ref{polar})
of the density matrix 
(cf. \cite{boya,bs}), the corresponding (conditional) Haar measure is simply 
$ \sin{\beta} d \beta d \alpha / 2 \pi$.
We converted the (generalized) ZHSL measures --- that is,
the product of this measure and  members of the (symmetric)
 Dirichlet family
(\ref{dirichlet}) --- to Cartesian coordinates, making use of the
transformations (several others, leading to equivalent results
(\ref{cartesian}) and (\ref{sph}), are also possible),
\begin{equation} \label{transformations}
\alpha = {1 \over 2} i (2 \log{(x + i y)} -\log{(x^2 +y^2)}),\quad
\beta= \cos^{-1}{(-{z \over \sqrt{x^2+y^2 +z^2}})},\quad
d_{1} = {1 \over 2} (1 - \sqrt{x^2 +y^2 +z^2}).
\end{equation}
By so doing, we obtained
the one-parameter ($\nu$) family of probability distributions,
\begin{equation} \label{cartesian}
q_{\nu}(x,y,z) = {\Gamma({1 \over 2} + \nu)(1- x^2 -y^2 -z^2)^{\nu-1}
\over  2 \pi^{{3 \over 2}} \Gamma(\nu) (x^2 +y^2 +z^2)},
\end{equation}
over the Bloch sphere.
(One can easily see, then, that the generalized ZHSL measures are
{\it  unitarily-invariant}, since the two 
eigenvalues of $\rho$ --- that is $(1 \pm \sqrt{x^2 +y^2 +z^2}) /  2$ --- are
preserved under unitary transformations. ``Our choice [of the ZHSL measure]
was motivated by the fact that both component measures are rotationally
invariant'' \cite{zycz2}.)
In spherical coordinates ($r, \theta, \phi$),
 this probability distribution (\ref{cartesian})
takes the form,
\begin{equation} \label{sph}
\tilde{q}_{\nu}(r,\theta,\phi) =
 {\Gamma({1 \over 2} + \nu) (1- r^2)^{\nu -1} \sin{\theta} \over
2 \pi^{3/2} \Gamma(\nu)}.
\end{equation}
Now, to cast (\ref{sph}) in the form 
(\ref{monotone}) of the volume element of a monotone metric 
(making the substitution $r = (1-t)/(1+t)$), one finds that
\begin{equation} \label{proportional}
f(t) \propto { (1-t)^2 \big( {t \over (1+t)^2}\big)^{1/2-\nu} \over (1+t)}.
\end{equation}
(Since the limit of these functions at $t =1$ for all $\nu$
 are 0, we are unable to 
normalize them so that $f(1) =1$, in the manner of \cite{petzsudar}, which 
condition would appear to be necessary for them to be
 associated with ``operator means''. However, 
interestingly, the functions
(\ref{proportional})
do fulfill the 
self-adjointness requirement of Petz and Sud\'ar \cite{petzsudar},
$f(t) = t f(1/t)$.)
 In particular, for $\nu =1$ (the uniform distribution on the 1-simplex
or line),
we obtain $f(t) \propto {(1-t)^2 \over \sqrt{t}}$, and for
$\nu = {1 \over 2}$ (that is, the Jeffreys' prior on the 1-simplex), we have
$f(t) \propto {(1-t)^2 \over (1+t)}$.
Now, the class of functions (\ref{proportional})
 possesses some interesting properties in
regard to monotonicity (as simple plots will reveal).
They are all monotone-{\it decreasing} for $t \in [0,1]$ (so they are
certainly not operator monotone in the sense of Petz and Sud\'ar 
\cite{petzsudar}), but they are also monotone-{\it increasing} for $t > 1$.
We do not know whether or not they are, then, also {\it operator} monotone
for $t > 1$ (but some of the 
numerous results  presented in \cite[Chap. V]{bhatia} might
be relevant in this regard). Nevertheless, it would appear that the behavior
of the functions given by (\ref{proportional}) in the interval [0,1] is 
possibly more
relevant than the behavior for $t > 1$, since in (\ref{monotone}), the
argument of $f$ --- that is, $(1-r)/(1+r)$ --- can vary only between 0 and 1.
(For the operator monotone functions $f(t)$, in the form 
considered by Petz and Sud\'ar \cite{petzsudar}, if one takes $y f(x/y)$, 
that is, the reciprocal of the associated ``Morozova-Chentsov function'', one
obtains various means of $x$ and $y$,
 such as the arithmetic and logarithmic ones \cite[eq. (3)]{petz}.
Using (\ref{proportional}) in this formula, one obtains for 
the case $\nu = 1$, $(x-y)^2 /4 \sqrt{x y}$,
and for $\nu ={1 \over 2}$, $(x-y)^2 /2 (x+y)$.))

We have, thus, shown  that the ``natural measure'' recently
proposed by ZHSL \cite{zycz1} 
(and a class of extensions of it) can not be considered (at least for the case
of two-dimensional quantum systems, but conjecturally also for 
the $N$-dimensional
systems, $N>2$, as well)
 to be proportional to the volume elements of  monotone metrics.
Any  metrics associable  with these ZHSL measures
would, therefore, lack the
statistically important  property of being
decreasing under stochastic mappings (that is, coarse-grainings).

Let us point out that we have previously, as well,  argued that the ZHSL
measure and its variants do not correspond to normalized volume elements
of monotone metrics \cite[secs. II.C, D]{slater3}. But there
 the  evidence 
(concerning the eigenvalues of certain averaged density matrices) was of a
more indirect nature.
In particular, we were unaware in \cite{slater3}
of the fact that it is possible to express (cf. \cite{boya,bs}) 
the ZHSL measure 
 using the theoretically minimal number 
($N^2-1$) of variables needed to parameterize the convex set of
$N \times N$ density matrices (rather than the naive number,
$N^2 +N -1$, that a reading of \cite{zycz1} would immediately suggest).
Consequently, in \cite{slater3} we did not 
utilize the transformations (\ref{transformations}) between the 
(economized number of) ZHSL
parameters and  more conventionally employed ones.

It would be of interest, as well, 
 to investigate to what extent it is possible to
replace the Dirichlet distributions (\ref{dirichlet}) in the 
ZHSL (product)  measure 
(leaving, however, intact the Haar measure)
 by other probability distributions lying outside the
Dirichlet family, so that the so-modified ZHSL 
(product) measures would, then in fact,
take the 
form of normalized volume elements of monotone metrics.
In fact, in \cite[eqs. (10), (11)]{buresprior} we have
 been able --- following the lead of
Hall \cite[eqs. (24), (25)]{hall} --- to do precisely
 this for the specific instance of the minimal monotone
(Bures) metric in the cases $N=2$ and 3.
Also in \cite{buresprior}, we have 
further determined the 
necessary Hall normalization constants (for the {\it marginal} 
Bures probability
distributions over the $N$ eigenvalues of the $N \times N$
density matrices) for 
$N=4$ and 5 which constants, in conjunction with Euler angle parameterizations 
(not yet fully specified) of
$SU(4)$ and $SU(5)$, parallel to that reported for
$SU(3)$ in \cite{byrd,byrdsudar}, would allow one to obtain the 
corresponding normalized Bures
volume elements for those $N$-dimensional quantum systems, as well.
The normalization constants ($N>2$) reported in 
\cite{buresprior} appear to be strongly related to partial sums of
the denominators of even-indexed Bernoulli numbers.
\

\acknowledgments

I would like to express appreciation to the Institute for Theoretical
Physics for computational support in this research, to Mark Byrd
for
his insightful observation that it is possible to express the ZHSL measure
in a non-``over-parameterized'' form, as well as to K. \.Zyczkowski for
his many informative communications.

\end{document}